\documentclass{aastex631}
\usepackage{CJKutf8}
\usepackage[caption=false]{subfig}
\usepackage{graphicx}
\usepackage{amsmath}
\usepackage{extarrows}
\usepackage{multirow}
\usepackage{graphicx}
\usepackage{booktabs}
\usepackage{amsmath,amssymb}
\usepackage[T1]{fontenc}
\graphicspath{{./}{figures/}}

\newcommand{\Ni}{{$^{56}$Ni }}
\newcommand{\Co}{{$^{56}$Co }}
\newcommand{\Fe}{{$^{56}$Fe }}
\newcommand{\Msun}{M$_\odot$}

\begin{document}

\title{The Energy Sources, the Physical Properties, and the Mass-loss History of SN~2017dio}

\correspondingauthor{Shan-Qin Wang \begin{CJK*}{UTF8}{gbsn}(王善钦)\end{CJK*}}
\email{shanqinwang@gxu.edu.cn}

\author{Deng-Wang Shi \begin{CJK*}{UTF8}{gbsn}(石登旺)\end{CJK*}}
\affiliation{Guangxi Key Laboratory for Relativistic Astrophysics,
	School of Physical Science and Technology, Guangxi University, Nanning 530004,
	China}

\author{Shan-Qin Wang \begin{CJK*}{UTF8}{gbsn}(王善钦)\end{CJK*}}
\affiliation{Guangxi Key Laboratory for Relativistic Astrophysics,
	School of Physical Science and Technology, Guangxi University, Nanning 530004,
	China}

\author{Wen-Pei Gan \begin{CJK*}{UTF8}{gbsn}(甘文沛)\end{CJK*}}
\affiliation{Nanjing Hopes Technology Co., Ltd. Nanjing, 210000, China}

\author{En-Wei Liang  \begin{CJK*}{UTF8}{gbsn}(梁恩维)\end{CJK*}}
\affiliation{Guangxi Key Laboratory for Relativistic Astrophysics,
	School of Physical Science and Technology, Guangxi University, Nanning 530004,
	China}

\begin{abstract}

We study the energy sources, the physical properties
of the ejecta and the circumstellar medium (CSM), as well as the mass-loss history of the progenitor
of SN~2017dio which is a broad-lined Ic (Ic-BL) supernova (SN) having unusual light curves (LCs)
and signatures of hydrogen-rich CSM in its early spectrum.
We find that the temperature of SN~2017dio began to increase linearly
about 20 days after the explosion. We use the $^{56}$Ni plus the ejecta-CSM interaction (CSI)
model to fit the LCs of SN~2017dio, finding that the masses of the ejecta, the $^{56}$Ni,
and the CSM are $\sim$ 12.41 M$_\odot$, $\sim$ 0.17 M$_\odot$, and $\sim$ 5.82 M$_\odot$, respectively.
The early-time photosphere velocity and the kinetic energy of the SN are respectively
{$\sim$ 1.89 $\times 10^4$ km s$^{-1}$} and $\sim$ 2.66 $\times 10^{52}$ erg, which are respectively comparable to those of
SNe Ic-BL and hypernovae (HNe).
We suggest that the CSM of SN~2017dio might be {from an luminous-blue-variable-like outburst or}
pulsational pair instability $\sim$ 1.2$-$11.4 yr prior to the SN explosion{, or binary mass transfer}.
{Moreover,} we find that its ejecta mass is larger than those of many SNe Ic-BL, and that
its $^{56}$Ni mass ($M_{\rm Ni}$) is approximately equal to the mean (or median) value
of $M_{\rm Ni}$ of SNe Ic-BL in the literature, but lower than $M_{\rm Ni}$ of
prototype HNe (e.g., SN~1998bw and SN~2003dh).

\end{abstract}

\section{Introduction}
\label{sec:introduction}

Core-collapse supernovae (CCSNe) are from the explosions of massive stars
whose Zero Age Main Sequence (ZAMS) masses are larger than 8 M$_\odot$.
The explosions leave neutron stars or black holes in their centers.
According to their spectra, most CCSNe can be
divided into types II, IIb, Ib, and Ic \citep{Filippenko1997}.
SNe II retain hydrogen envelopes before explosions;
SNe Ib lost hydrogen envelopes before explosions; SNe Ic lost hydrogen and almost all helium envelopes before explosions \citep{Filippenko1997}.
A minor fraction of SNe Ic are broad-lined ones (SNe Ic-BL) whose absorption lines in spectra are broader than those of
normal SNe Ic. The derived photospheric velocities of many SNe Ic-BL are $\gtrsim$ {$2\times10^4$ km s$^{-1}$} (see, e.g., \citealt{Woosley2006}, \citealt{Hjorth2012}
, and references therein),
which are significantly higher than those of normal SNe Ic \citep{Woosley2006}.
Moreover, some SNe Ic-BL have kinetic energy ($E_{\rm K}$) $\gtrsim$ $10^{52}$ erg, becoming
hypernovae (HNe, e.g., \citealt{Iwamoto1998}; \citealt{Iwamoto2000}). \footnote{\cite{Pac1998} propose the term "hypernova" to name the explosive phenomenon (especially for the luminous optical
afterglows associated with observed gamma-ray bursts) that is much more luminous and energetic than any SN.
Following studies modeling SNe Ic-BL (e.g., \citealt{Iwamoto1998}; \citealt{Iwamoto2000})
use "hypernova" to name the SNe that are much more energetic than normal SNe which have kinetic energy $\sim 10^{51}$ erg.
Most HNe have kinetic energy $\gtrsim$ $10^{52}$ erg (see, e.g., \citealt{Iwamoto1998}; \citealt{Iwamoto2000}, Table 9.2 of \citealt{Hjorth2012}
and references therein).} It is widely believed that the main energy of the light curves (LCs) of most
SNe Ib and Ic (including SNe Ic-BL and HNe) are from the cascade decay of the synthesized $^{56}$Ni by the SNe
(e.g., \citealt{Arnett1982,Cappellaro1997,Valenti2008,Drout2011,Cano2013,Taddia2015,Wheeler2015,Lyman2016,Prentice2016,Taddia2019}).

In the past four decades, a few hundred interacting CCSNe \citep{Smith2017}
with {narrow and intermediate-width} emission lines indicative of the
interaction between the ejecta and circumstellar medium (CSM) have been confirmed
(e.g., \citealt{Schlegel1990,Mattila2008,Pas2008a,Pas2008b,Pastorello2016,2021arXiv210807278F,Perley2022,Gal-Yam2022,Davis2023}).
{It is believed that the emission lines are produced by the ejecta-CSM interaction (CSI) \citep{Schlegel1990,Filippenko1997}.}
According to the features of their spectra, interacting CCSNe can be divided into types IIn \citep{Schlegel1990},
Ibn \citep{Mattila2008,Pas2008a,Pas2008b,Pastorello2016}, and
Icn \citep{2021arXiv210807278F,Perley2022,Gal-Yam2022,Davis2023}.
The three subclasses respectively show hydrogen emission lines, helium emission lines,
and helium-deficient emission lines, indicating the existence of hydrogen-rich (for SNe IIn),
helium-rich (for SNe Ibn), and helium-deficient (for SNe Icn) CSM.

The forward and reverse shocks {produced by the CSI}
accelerate the electrons. Accelerated electrons emit high-energy ($\gamma$-ray and X-ray)
photons, which can be thermalized to be UV-optical-IR photons
if the densities of CSM are large enough. The emission produced by
the CSI can provide additional or main energy source for powering the LCs
of the interacting SNe \citep{Chevalier1982,ChevalierFransson1994}.

SN~2017dio is a peculiar interacting SN \citep{Kuncarayakti2018} which was
discovered on April 26, 2017 and then confirmed to be an SN Ic in the
galaxy SDSS J113627.76+181747.3 whose redshift $z$ is 0.037 (see \citealt{Kuncarayakti2018} and references therein).
The follow-up observations performed by \cite{Kuncarayakti2018} obtain the optical ($ugriz$) and {near-infrared (NIR)} ($JHK$) LCs of SN~2017dio.
The early-time $griz$ LCs of SN~2017dio are featured by the rising episodes with dips or short plateaus, while the late-time
LCs are {slow-evolving second peaks} (see Figure 2 of \citealt{Kuncarayakti2018}).
Based on the analysis for spectra, \cite{Kuncarayakti2018} find that SN~2017dio is an SN Ic-BL showing H$\alpha$ emission lines
, and conclude that it is the first confirmed SN Ic having signatures of hydrogen-rich CSM in the early spectrum.
The feature of the spectral evolution of SN~2017dio is different to that of normal SN Ic (including SN Ic-BL)
whose spectra do not show CSI signatures and that of SNe Icn which collide with hydrogen and helium-deficient CSM.
On the other hand, the main features of SN~2017dio are reminiscent of SN~2017ens which is a superluminous SN (SLSN) showing the
transition from an SN Ic-BL to an SN IIn \citep{Chen2018}.

The early-time LCs of SN~2017dio are well consistent with the template LCs of SNe Ic \citep{Kuncarayakti2018},
indicating that they might be powered by \Ni cascade decay; the rebrighening of the LCs and the H$\alpha$
emission lines in the spectra favor the scenario that the late-time LCs are mainly powered
by CSI between the ejecta of SN~2017dio and the hydrogen-rich CSM surrounding the
SN progenitor \citep{Kuncarayakti2018}.

The observations and detailed modeling for the constructed
bolometric LC of SN~2017ens are performed by \cite{Chen2018}. For comparison,
the {physical properties} of the ejecta, the CSM, and the mass-loss history of the progenitor of
SN~2017dio have not been quantitatively constrained, though
the observations and some excellent analyses are performed by \cite{Kuncarayakti2018}.

We suggest that the physical properties of SN~2017dio and the CSM surrounding the SN deserve further study,
since the detailed modeling would reveal the energy sources, physical properties of the SN ejecta and the CSM,
as well as the mass-loss history of the SN progenitor. Hence, we perform a detailed modeling for its multi-band LCs.
In Section \ref{sec:LCfits}, we constrain the physical properties of SN~2017dio as well as the CSM
by using the \Ni plus CSI model to fit its multi-band LCs.
In Section \ref{sec:dis}, we {discuss the possible mechanism of the mass loss,} infer the mass-loss history of the progenitor,
and compare the physical properties of SN~2017dio to those of
SN~2017ens and other SNe Ic-BL.
In Section \ref{sec:con}, we draw some conclusions.
Throughout the paper, we assume $\Omega_m = 0.315$,
$\Omega_\Lambda = 0.685$, and $H_0 = 67.3$\,km\,s$^{-1}$\,Mpc$^{-1}$ \citep{Planck2014}. The values of the Milky Way
reddening ($E_\mathrm{B-V}$) are from \cite{Schlafly2011}.

\section{Modeling the Multi-band Light Curves of SN~2017dio Using the \Ni Plus CSI Model}
\label{sec:LCfits}

As pointed out by \cite{Kuncarayakti2018}, the early-time LCs which include the first peaks of SN~2017dio resemble those of
SNe Ic (see template match using SNe Ic LCs in Figure 2 of \citealt{Kuncarayakti2018}) which are powered by \Ni cascade decay,
while the rebrightening LCs and prominent hydrogen and helium emission features in the spectra at the
epochs after the first peaks suggest that the {second peaks} of the LCs of SN~2017dio might be powered by \Ni cascade decay plus CSI.

Hence, we use the \Ni plus CSI model to fit the LCs of SN~2017dio and
assume the onset time of the CSI is later than that of \Ni cascade decay.
The time interval is set to be the difference between the onset time of CSI
and the time of the first photometric point ($\Delta t$) minus the explosion time
relative to the first data ($t_{\rm shift}$), i.e., $\Delta t-t_{\rm shift}$.

The \Ni model we use to reproduce the bolometric LC is from equations (1) and (5)
of \cite{WangTao2023}, which is based mainly on \cite{Arnett1982} and \cite{Valenti2008};
\footnote{\cite{Arnett1982} don't consider energy from the process of \Co to \Fe and the leakage effect of $\gamma$-ray and positions;
\cite{Valenti2008} include the \Co to \Fe process and the leakage effect.}
the equations of the CSI model used here can be found in \cite{Wang2019}, which is based on
\cite{Chevalier1982}, \cite{ChevalierFransson1994}, and \cite{Chatzopoulos2012}.
\footnote{\cite{Wang2019} assume stationary photospheres at the early-time epochs, while
we assume that the photosphere expanded at the early-time epochs. Both cases are considered in
Section 2 of \cite{Chatzopoulos2012}.}

The density of CSM can be written as
\begin{eqnarray}
\rho_{\rm CSM}=q r^{-s}
\end{eqnarray}
where \begin{eqnarray}
q=\rho_{\rm CSM,1}r_1^{s}
\end{eqnarray}
here, $r_1=v_{\rm ph}\left(\Delta t-t_{\rm shift}\right)$ is the innermost radius of the CSM, $\rho_{\rm CSM,1}$ is the density
of the CSM at $r_1$. The power law index ($s$) is assumed to be 0, since \cite{Kuncarayakti2018} suggest that the CSM might be a shell whose density
is constant at different radii; therefore, $\rho_{\rm CSM}=\rho_{\rm CSM,1}$. The optical opacity ($\kappa$) of the hydrogen-deficient ejecta and the hydrogen-rich CSM are set
to be 0.07 (e.g., \citealt{Taddia2015}) and 0.34 cm$^2$~g$^{-1}$, respectively;
the $\gamma$-ray opacity ($\kappa_\gamma$) and the ${e}^{+}$ opacity ($\kappa_{e^{+}}$) are fixed to be 0.027 cm$^2$~g$^{-1}$ (e.g., \citealt{Cappellaro1997,Mazzali2000,Maeda2003}) and 7 cm$^2$~g$^{-1}$ (\citealt{Cappellaro1997}), respectively.

\begin{figure}[htbp]
	\centering
	\includegraphics[width=0.48\textwidth,angle=0]{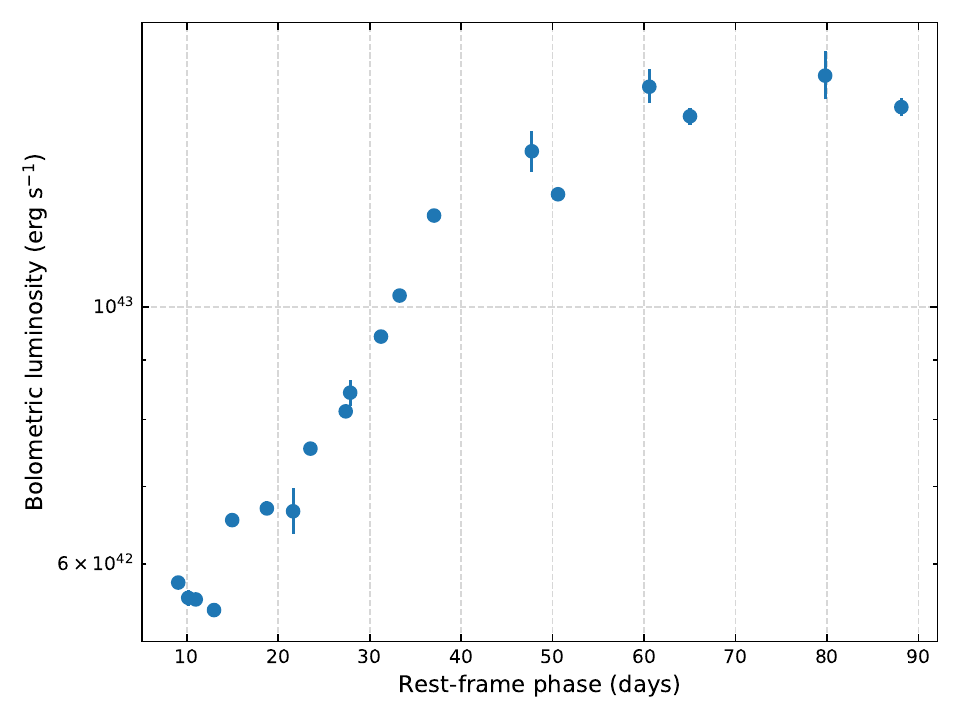}
	\includegraphics[width=0.48\textwidth,angle=0]{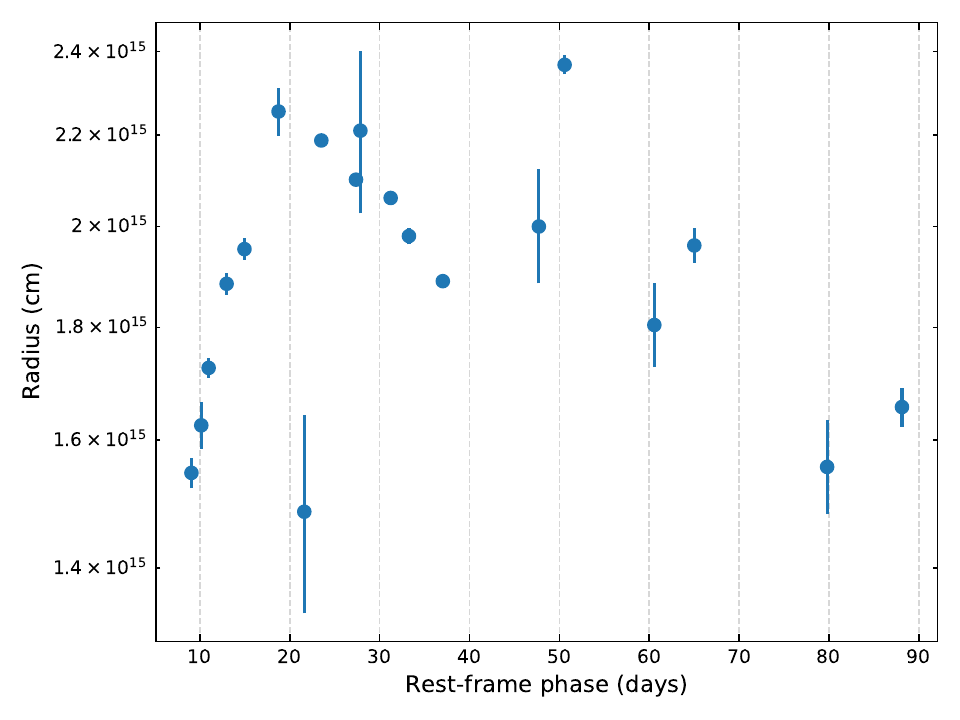}
	\includegraphics[width=0.48\textwidth,angle=0]{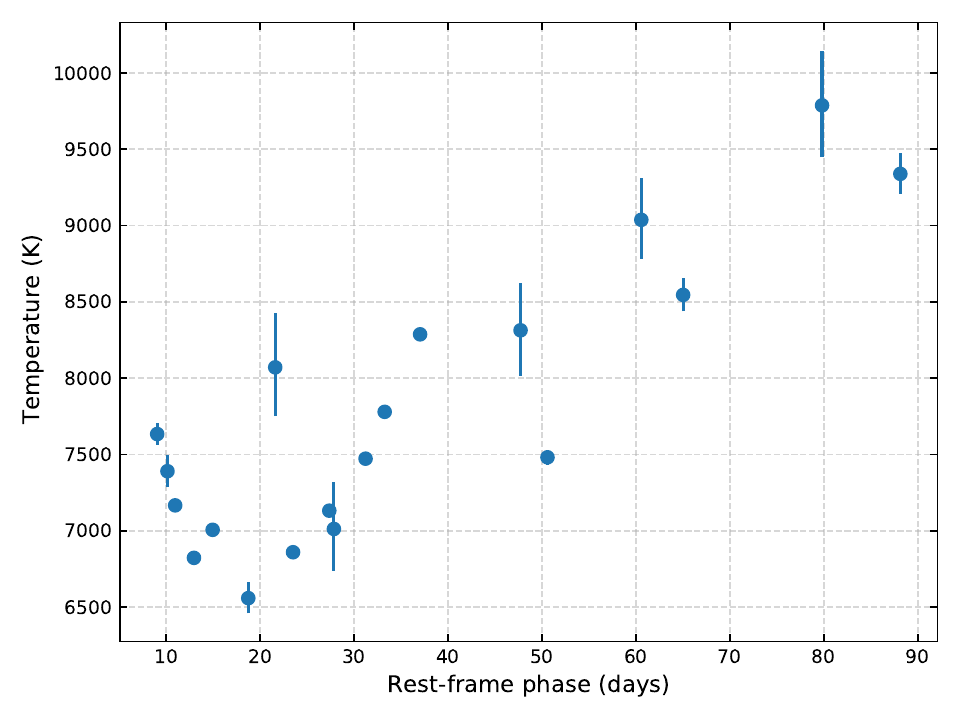}
	\caption{The bolometric luminosity, blackbody radius and blackbody temperature of SN~2017dio derived by fitting its SEDs at different epochs.}
	\label{fig:LRT}
\end{figure}

To fit the multi-band LCs, a photosphere modulus is needed. By using the blackbody model
($F_{\nu, \mathrm{ph}}= \frac{2{\pi}h \nu^{3}}{c^{2}}\left(e^{\frac{h \nu}{k_{b} T_{\mathrm{ph}}}}-1\right)^{-1} \frac{R_{\mathrm{ph}}^{2}}{D_{L}^{2}}$,
where $R_{\mathrm{ph}}$ and $T_{\mathrm{ph}}$ are the radius and temperature of SN photosphere,
$D_{L}$ is the SN luminosity distance) to fit the optical-NIR SEDs at all epochs
having photometry in at least three bands (see Figure \ref{fig:sed},
the corresponding best-fitting parameters are listed in Table \ref{table:blackbody}),
we find that the derived bolometric LC {(the top-left panel of Figure \ref{fig:LRT})} shows a dip (days $\sim$ 9 to $\sim$ 15 after the time of the first photometric data) and
a {slow-evolving second peak} (days $\sim$ 60 to $\sim$ 90), which are consistent to the features of the multi-band LCs.
Additionally, it can be found that the early-time velocity of the photosphere of the SN {wa}s approximately constant (the top-right panel of Figure \ref{fig:LRT}),
while the temperature of the SN increase{d} linearly about 20 days after the explosion (the bottom panel of Figure \ref{fig:LRT}).

Therefore, we assume that the evolution of photosphere radius
can be described by Equation (9) of \cite{Nicholl2017}, while
the temperature evolution can be described the {equation} below
\begin{eqnarray}
T_{\rm ph}(t) =
\left\{
\begin{array}{lr}
	\left(\frac{L_{\rm SN}(t)}{4 \pi \sigma v_{\rm ph}^2 t^2}\right)^{\frac{1}{4}},\ &
	\qquad t\le t_b \\
	\alpha_{\rm pb}(t-t_b)+T_b,&
	\qquad t > t_b \\
\end{array}
\right.
\label{equ:T_ph}
\end{eqnarray}
$T_{\rm b}$ is the lowest photosphere temperature.
This introduces two new free parameters, $t_{\rm b}$ which is the time when the temperature began to increase and $\alpha_{\rm pb}$
which is the slope of the temperature curve after $t_{\rm b}$.

The definitions, the units, and the prior ranges of the parameters of the \Ni plus CSI model are shown in Table \ref{table:NI+CSI}.
The Markov Chain Monte Carlo (MCMC) using the \texttt{emcee} Python package \citep{Foreman-Mackey2013} is adopted to get the best-fitting parameters and the 1-$\sigma$ {bounds}
(which are corresponding to the 16th and 84th percentiles of the posterior samples) of the parameters.

\begin{figure}[htbp]
	\centering
	\includegraphics[width=0.7\textwidth,angle=0]{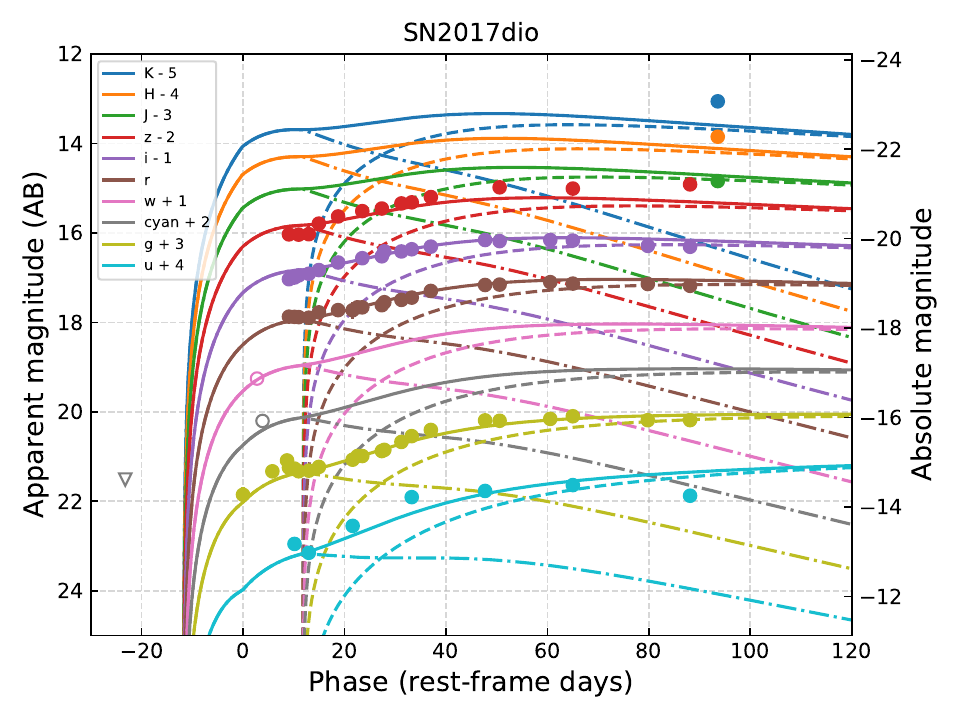}
	\caption{The best fits (the solid curves) of the multi-band LCs
		of SN~2017dio using the \Ni plus CSI model. The open triangle represents the $cyan$-band upper limit, the open circles represent the data of ATLAS and PS1 ($cyan$- and $w$- bands), the filled circles
		represent the data in $ugriz$ and $JHK$ bands which are from \cite{Kuncarayakti2018}. The dash-dot and dashed lines represent the contributions from \Ni decay and CSI, respectively.
		The shaded regions indicate 1-$\sigma$ bounds
		of the parameters.}
	\label{fig:fitting}
\end{figure}

All available photometric data which are in $ugriz$ and $JHK$ bands are from \cite{Kuncarayakti2018},
except for one upper limit ($cyan$-band) and two data points ($cyan$- and $w$- bands) which are from the Asteroid Terrestrial-impact Last Alert System (ATLAS; \citealt{Tonry2011})
and the Panoramic Survey Telescope and Rapid Response System (Pan-STARRS, PS1, \citealt{Flewelling2020}).
\footnote{https://www.wis-tns.org/object/2017dio}
The fitting of the LCs of SN~2017dio using the \Ni plus CSI model is shown in Figure \ref{fig:fitting},
the best-fitting parameters are listed in Table \ref{table:NI+CSI},
the corresponding corner plot is presented in Figure \ref{fig:corner}.
It can be found that the first peaks of SN~2017dio {can be} fitted by \Ni model;
the LCs would decline after the first peaks
(see the dash-dotted lines of Figure \ref{fig:fitting}) if the CSI was absent.
The contribution of the CSI between the ejecta and the CSM (see the dashed lines of Figure \ref{fig:fitting})
results in the rebrightening of the LCs.

The last optical-NIR photometry (at 88.13 day) cannot be fitted by the model
(the $u-$ and $z-$band data deviate from the theoretical LC),
since the optical-NIR SED at this epoch deviates from blackbody model
(see the last panel of Figure \ref{fig:sed}). The photometry at the
last epoch when only NIR data is available also can not be fitted by the
model; the $H-$ and $K-$band flux is higher than the theoretical
LC, indicating that there might be NIR excess at this epoch.

The best-fitting parameters of the \Ni plus CSI model are
$M_{\rm ej}$ = 12.41 M$_\odot$,
{$v_{\rm ph}$ = 1.89 $\times 10^4$ km s$^{-1}$},
$M_{\rm Ni}$= 0.41 M$_\odot$,
$M_{\rm CSM}$ = 5.82 M$_\odot$,
$\rho_{\rm CSM}$ = 3.16  $\times10^{-15}$ g cm$^{-3}$,
$\Delta t$ = 11.84 days,
$\epsilon$ = 0.28,
$x_{\rm 0}$ = 0.28,
$t_{\rm b}$ = 11.57 days,
$\alpha_{\rm pb}$ = 41.50 K d$^{-1}$,
$t_{\rm shift} = $ $-11.74$ days.
The values of reduced $\chi^2$ ($\chi^2$/dof, dof is the degree of freedom) is 114.48.

It should be noted that, however, \cite{Arnett1982}'s model and {it's generalized versions} (including {the one} we adopt) would overestimate the
\Ni masses of stripped{-}envelope SNe (including SNe Ic and Ic-BL) (see, e.g., \citealt{Khatami2019}).
Thus we calculate {more} accurate value of \Ni mass of SN~2017dio
using the equation below \citep{Khatami2019}
\begin{eqnarray}
M_{\rm Ni} = \frac{L_{\rm p} \beta^2 t_{\rm p}^2 }{2 \epsilon_{\rm Ni} \tau_{\rm Ni}^2} \left( \left(1 - \frac{\epsilon_{\rm Co}}{\epsilon_{\rm Ni}} \right)\times (1 -
(1 + \beta t_{\rm p}/\tau_{\rm Ni})e^{-\beta t_{\rm p}/\tau_{\rm Ni}}) + \frac{\epsilon_{\rm Co} \tau_{\rm Co}^2}{\epsilon_{\rm Ni} \tau_{\rm Ni}^2} \left(1 - (1+\beta t_{\rm p}/\tau_{\rm Co} )e^{-\beta t_{\rm p}/\tau_{\rm Co}} \right)  \right)^{-1}
\label{equ:newNimass}
\end{eqnarray}
The value of $\beta$ is $\sim$0.56 for SNe Ic-BL \citep{Afsariardchi2021}.
Using the peak luminosity ($L_{\rm p}$) and the rise time ($t_{\rm p}$)
of the bolometric LC produced by the best-fitting parameters of the multi-band
LC fit, we find that the $M_{\rm Ni}$ of SN~2017dio is 0.17 M$_\odot$.

{Assuming} that the CSI was triggered when the LCs began to rebrighen,
the time interval {($\Delta t_1$)} from SN explosion to the time when CSI was triggered {($\Delta t_1=\Delta t-t_{\rm shift}$;}
$\Delta t=11.84$ days, $t_{\rm shift}=-11.74$ days, see Table \ref{table:NI+CSI})
is 23.58 days. This value is about 1/3 times the value ($\sim$ 80 days) inferred by \cite{Kuncarayakti2018} {which
assume that the CSI was triggered when H$\alpha$ luminosity reached its peak}.
Using our derived photometric velocity ($1.89 \times 10^4$ km s$^{-1}$) which is about twice that adopted by \cite{Kuncarayakti2018}
($10^4$ km s$^{-1}$) \footnote{{The $v_{\rm ph}$ value of \cite{Kuncarayakti2018} is from the mean value of SNe Ibc in \cite{Cano2013}.
Note that the average $v_{\rm ph}$ values of SNe Ib, Ic, and SNe Ic-BL are $\sim 8000 \pm 1700$ km s$^{-1}$,
$\sim 8500 \pm 1700$ km s$^{-1}$, and $\sim 15000 \pm 4000$ km s$^{-1}$, respectively (see, e.g., Table D1 of \citealt{Cano2013}).
We suggest that the mean value of $v_{\rm ph}$ of SNe Ic-BL is more reasonable for SN~2017dio, since it is a SN Ic-BL.}},
the derived value of the innermost radius of CSM $r_1$ ($r_1=v_{\rm ph}\Delta t_1$) is
$3.85\times 10^{15}$ cm (257 au), which is about half of the inferred value of \cite{Kuncarayakti2018} ($\sim$ 500 au).

Combining the equation $M_{\rm CSM} = \frac{4}{3}\pi \rho_{\rm CSM}(r_2^3-r_1^3)$ and the derived values of $\rho_{\rm CSM}$, $r_{1}$, {and}
$M_{\rm CSM}$, we find that the outermost radius $r_{\rm 2}$ and the thickness { ($\Delta{r}$, $\Delta{r}=r_2-r_1$) of the CSM}
are 9.37 $\times 10^{15}$ cm (625 au) and 5.52 $\times 10^{15}$ cm (368 au), respectively.

\section{Discussion}
\label{sec:dis}

\subsection{The {Mechanism and History of the Mass loss of} the Progenitor}
\label{subsec:mass-loss}

\cite{Kuncarayakti2018} exclude the possibility that the CSM of SN~2017dio is a
steady-state stellar wind, and favor the scenario that the CSM is a shell expelled
by the SN progenitor itself or from the binary companion.

{The eruptions of massive stars can be triggered by the luminous-blue-variable (LBV)-like outbursts \citep{Smith2011} or}
the pulsational pair instability (PPI) \citep{Woosley2007,Woosley2017,Marchant2019,Renzo2020}.

{For the LBV-like outburst scenario, t}he derived CSM mass ($\sim$ 5.82 M$_\odot$) {is in the range of
the masses of CSM from LBV-like outbursts ($\sim$0.01 to $\sim$10 \Msun, \citealt{Smith2011}).
Assuming that the compact remnant is a neutron star with mass $\sim 1.4$ \Msun or
a stellar mass black hole, the mass of the progenitor of SN~2017dio is $\gtrsim$ 19.6 \Msun ($M_{\rm ej}=12.41$ \Msun).
The ZAMS mass of the progenitor must be larger than this value, because the wind can reduce the mass of the progenitor.
\cite{Smith2011} find that the initial (i.e., ZAMS) masses of some SN impostors which are related to
the eruptions of massive hydrogen-rich stars are in the range of $10-20$ \Msun, which is significantly
lower than the expected the range of ZAMS masses that can experience violent eruptions.
This suggests that the ZAMS masses of massive stars that can trigger LBV-like eruptions
can be down to 10 \Msun or even below 8 \Msun \citep{Smith2011}.
Our derived progenitor mass of SN~2017dio ($\gtrsim$ 19.6 \Msun) indicates that it can
experience LBV-like eruption.}

{For the PPI scenario, the derived CSM mass} is comparable with the values derived in the literature
(see, e.g., Table 2 of \citealt{Liu2018}, Extended Data Table 1 of \citealt{Lin2023})
using {this scenario}.
{The fact that the CSM of SN~2017dio is hydrogen-rich suggests that the metallicity ($Z$) of the progenitor of SN~2017dio
might be lower than those of the progenitors of type I SNe (e.g., SN~2017egm) whose progenitors are assumed to experience PPI, since it retained
a moderate amount of hydrogen envelope prior to the PPI.
\footnote{{The wind mass-loss rate $\dot{M}$ of stars is proportional
$Z^m$ \citep{Kudritzki1987,Leitherer1992,Vink2001A&A,Mokiem2007}, $m = 0.69\pm0.10$ \citep{Vink2001A&A}
or $0.83\pm0.16$ \citep{Mokiem2007}. It is more difficult to deplete the hydrogen envelopes via winds for low $Z$ stars.}}
Table 2 of \cite{Woosley2017} lists the low {metallicity} ($Z=0.1\,Z_\odot$) PPI models.
{For T80, T80A, T80B, T80C, and T80D Models in which the outbursts might correspond to type IIn SNe or SN impostors \citep{Woosley2017}
,} the masses of the hydrogen envelopes of the pre-SN cores are in the range of {16.09$-$43.6} \Msun;
it can be expected that {a higher} $Z$ would result in {a lower} hydrogen envelope {mass} before the PPI.
our derived CSM mass might be reasonable, if the progenitor had a higher $Z$ and/or experienced more than one outburst.
In this scenario, the {mass of the remnant is $\lesssim 22.18-23.99$ \Msun},
since the mass{es} of the pre-SN {helium} core{s of T80, T80A, T80B, T80C, and T80D Models
which can be set to be the upper limit of the progenitor mass just before the
core collapse are in the range of} {$34.59-36.4$ \Msun} (see Table 2 of \citealt{Woosley2017}),
while the ejecta mass is $\sim 12.41$ \Msun.} {On the other hand, the lower limit of mass of the remnant is
$\sim 16.36-18.17$ \Msun which corresponds to the extreme case that all the 5.82 \Msun of CSM is helium.
Therefore, the remnant might be a black hole with mass in the range of $\sim 16.36-23.99$ \Msun.}

We can infer the mass-loss history of the progenitor of SN~2017dio,
if the CSM was expelled by an eruption of the progenitor shortly before the SN explosion.
{\cite{Smith2011} find that the velocities of hydrogen-rich CSM associated with SN impostors
are in the range of $100-1000$ km s$^{-1}$ (see also Table 9 of \citealt{Smith2011}). On the other hand,
\cite{Woosley2017} point{s} out that the typical velocity of the shells expelled by PPI {are 200-800} km s$^{-1}$ for Model {T80A
and 300-1000 km s$^{-1}$ for Models T80B, T80C, and T80D}; it is reasonable to assume that the velocity of the shell expelled by
PPI is $\sim$ {200-}1000 km s$^{-1}$. Hence, we assume that $v_{\rm shell}$ is $\sim 100-1000$ km s$^{-1}$.}
Using $v_{\rm ph}\Delta t_1 = v_{\rm shell}(\Delta t_0+\Delta t_1)$ ($\Delta t_0$ is {the} time interval
from the shell expelled to SN explosion,), we find the value of $\Delta t_0$ is $\sim 422-4174$ days ($\sim 1.2-11.4$ yr).
Therefore, the progenitor of SN~2017dio might experience an eruption $\sim$ 1.2$-$11.4 yr prior to the SN explosion.

{For {the} binary star scenario, the mass transfer rate $\dot{M}$ of massive stars via Roche-lobe overflow (RLOF)
can be of order $\sim 10^{-3}$ \Msun yr $^{-1}$ or higher \citep{Taam2000,Langer2012,Smith2014},
while the thermal (Kelvin-Helmholtz) timescale $\tau_{\rm KH}$ of the envelope is $\sim 10^{4}$ yr (\citealt{Smith2014},
the value of $\tau_{\rm KH}$ can be estimated by using Equation 10 of \citealt{Pac1971}).
Using $\Delta{M}=\dot{M}\tau_{\rm KH}$, the mass transferred can be up to $\sim 10$ \Msun or higher.
Our derived CSM mass value is in the range of the mass transferred in binary systems.}

\subsection{Comparison With SN~2017ens}
\label{subsec:Comparison-17ens}

The main features of SN~2017dio resemble those of SN~2017ens which is an SLSN Ic. The {spectra of SN~2017ens
at $\sim$ 1 month after peak show some broad features which resemble those of SNe Ic-BL at similar epochs},
while its late-time spectra indicate that it became an SN IIn \citep{Chen2018}.
Although {the} spectral evolution of {SN~2017dio and SN~2017ens} share similar features, {they also}
show some discrepancies.

(1). The bolometric luminosity of the first peak of SN~2017dio ($\sim 5.6 \times 10^{42}$ erg s$^{-1}$)
is lower than that of its {second peak} ($\sim 1.5 \times 10^{43}$ erg s$^{-1}$),
while the bolometric luminosity of the peak of SN~2017ens ($\sim 5.86\times 10^{43}$ erg s$^{-1}$, \citealt{Chen2018})
is higher than that of its late-time plateau ($\sim 3.0 \times 10^{42}$ erg s$^{-1}$,
see the bottom panel of Figure 1 of \citealt{Chen2018}).

(2). The peak bolometric luminosity of SN~2017ens is $\sim$10 times that of the bolometric luminosity of the first peak of SN~2017dio.
\cite{Chen2018} demonstrate that the \Ni model cannot account for the first peak of SN~2017ens, but the magnetar/fallback plus CSI model
can explain it. In contrast, our modeling show{s} that the \Ni model can explain the first peak of SN~2017dio.
Without a central engine (magnetar or fallback) or CSI, the peak {luminosity} of SN~2017ens might be {comparable to that of} the first peak of SN~2017dio.

(3). The bolometric luminosity of plateau of the SN~2017ens is $\sim$ 1/5 times that of
the {second peak} of SN~2017dio, indicating that the contribution of CSI between the ejecta of SN~2017dio and its CSM
is higher than that of SN~2017ens.

(4). Based on the scenario that the {early-time} peak of SN~2017ens was powered mainly by a magnetar or fallback,
\cite{Chen2018} suggest that the innermost radius of CSM of SN~2017ens is 1.2 $\times 10^{15} $cm.
This value is $\sim$1/3 times that of SN~2017dio (3.85 $\times 10 ^{15} $cm).

\subsection{Comparison With Other Broad-lined Type Ic SNe and HNe}
\label{subsec:Comparison-HNe}

Based on the best-fitting parameters derived, we can compare the physical properties
of SN~2017dio to several prototype SNe Ic-BL and HNe whose physical properties had been comprehensively studied {in} the literature.
The main parameters of the SNe Ic-BL and HNe are listed in Table \ref{table:BL-Lc}.

The derived photospheric velocity ($v_{\rm ph}$) of SN~2017dio is {1.89 $\times 10^4$ km s$^{-1}$},
which is comparable to other SNe Ic-BL (e.g., SN~1997ef, SN~2006aj{,} see Table \ref{table:BL-Lc}),
and favored by \cite{Kuncarayakti2018}'s conclusion that SN~2017dio is an SN Ic-BL.
The ejecta mass of SN~2017dio (12.41 M$_\odot$) is larger than those of {HNe in Table \ref{table:BL-Lc}}.
{In the single-star framework, this might indicate that the progenitor of SN~2017dio
had a ZAMS mass larger than those of HNe or {a} lower {metallicity} \footnote{{Having the same ZAMS masses, single stars with lower
{metallicity} would result in lower $\dot{M}$ and larger pre-SN masses, see footnote 6 for details.}}, or produced a lighter compact remnant.
In the binary framework, a higher ejecta mass might
suggest that less material of the progenitor was accreted by the companion, and {a} more massive pre-SN
core was left just prior to the SN explosion.}

Using the equation $E_{\rm K} = 0.3M_{\rm ej}v_{\rm ph}^{2}$, we find
$E_{\rm K}$ of SN~2017dio is 2.66 $\times 10^{52}$ erg. This indicates that SN~2017dio might also be an HN.
The value of $E_{\rm K}$ of SN~2017dio is comparable to those of prototype HNe (e.g., SN~1998bw and SN~2003dh),
and higher than that of SN~1997ef which is also an HN (see Table \ref{table:BL-Lc}).

The value of $M_{\rm Ni}$ of SN~2017dio is smaller than
those of SN~1998bw and SN~2003dh, but comparable to that of
SN~1997ef (see Table \ref{table:BL-Lc}).
Moreover, the \Ni mass (0.17 M$_\odot$) of SN~2017dio is approximately equal to
the mean and median values (0.15 M$_\odot$) of SNe Ic-BL reinvestigated by \cite{Afsariardchi2021}.
The value is also smaller than the median value (0.26 M$_\odot$) of
SNe Ic-BL studied by \cite{Cano2013} as well as the median value (0.292 M$_\odot$) of
SNe Ic-BL in the {Palomar Transient Factory (PTF; \citealt{Rau2009}; \citealt{Law2009})} sample studied by \cite{Taddia2019}.
As mentioned above, however, \cite{Arnett1982}'s model which was also used by \cite{Cano2013} and \cite{Taddia2019}
would overestimate the $M_{\rm Ni}$ of stripped-envelope SNe (including SNe Ic-BL); using
equation \ref{equ:newNimass}, the medians of $M_{\rm Ni}$ of SNe Ic-BL can be reduced by a factor
of $\sim$ 0.5, and comparable to those of \cite{Afsariardchi2021} and the derived $M_{\rm Ni}$ of SN~2017dio.

\section{Conclusions}
\label{sec:con}

SN~2017dio is an SN Ic-BL that had been confirmed to interact with CSM by its spectral evolution
and the LC rebrightening. In this paper, we study the energy sources as well as the physical properties
of the SN and the CSM surrounding the SN progenitor.

We use the \Ni plus CSI model and blackbody model to fit the multi-band
LCs of SN~2017dio. A new photospheric modulus assuming a linearly increasing
temperature evolution is adopted since the fitting for the optical-NIR SEDs at different epochs
demonstrate that its temperature evolution linearly increase{d} about 20 days after the explosion.

We find that the ejecta mass, the \Ni mass, and the CSM mass are $\sim$ 12.41 M$_\odot$,
$\sim$ 0.17 M$_\odot$, and $\sim$ 5.82 M$_\odot$, respectively.
The early-time photosphere velocity and the kinetic energy of the SN are respectively
$\sim$ {1.89 $\times 10^4$ km s$^{-1}$} which is comparable to those of SNe Ic-BL
and $\sim$ 2.66 $\times 10^{52}$ erg which is comparable to those of HNe.
Therefore, SN~2017dio might be an HN embedded in a hydrogen-rich CSM.
By comparing the derived parameters of SN~2017dio to those of
SNe Ic-BL and HNe in the literature, we find that the ejecta mass of SN~2017dio is larger than those of many SNe Ic-BL and HNe;
the derived \Ni mass of SN~2017dio is approximately equal to the mean (or median) value
of those of SNe Ic-BL and SN~1997ef which is an HN, but lower than the \Ni masses of prototype HNe (SN~1998bw and SN~2003dh).

{The CSM can be from} an eruption of the progenitor {of SN~2017dio or the companion of the progenitor via the mass transfer process.}
{In the single-star framework}, we find that the progenitor experience an eruption expelling a shell $\sim$ 1.2$-$11.4 yr prior to the SN explosion,
{if the eruption was caused by an LBV-like outburst or PPI}.
The derived CSM mass is comparable with the values derived in the literature {adopting the two scenarios},
indicating that the CSM of SN~2017dio might be expelled by an eruption caused by {an LBV-like outburst or} PPI.
{In the binary framework, the derived CSM mass is lower than the upper limit ($\gtrsim$ 10 \Msun)
of the transferred mass in binary systems.}

By comparing the LCs and energy sources of SN~2017dio to those of SLSN SN~2017ens, we find that
SN~2017dio and SN~2017ens show some differences, though the spectral evolution of the two SNe are similar.
The most prominent difference is that the peak luminosity of SN~2017ens, which is $\sim$10 times that of the first peak of SN~2017dio,
cannot be explained by the \Ni model; a central engine (a magnetar or fallback) or CSI is needed
to account for the high luminosity of the first peak of SN~2017ens. For comparison, the first peak of SN~2017dio can be well
explained by the \Ni model. We suggest that the peak of SN~2017ens might be comparable to that of
the first peak of SN~2017dio, if the putative central engine (magnetar or fallback) or CSI was absent.
In other words, the first peak of SN~2017dio can reach significantly higher luminosity if it was boosted by a central engine.

To date, SNe with LCs and spectral evolution like SN~2017dio and SN~2017ens are very rare.
The upcoming sky surveys for SNe, follow-up observations as well as detailed modeling
can shed light on the nature of them.

\clearpage

\begin{acknowledgements}
{We thank the anonymous referee for helpful comments and suggestions that have allowed us to improve this manuscript.}	
Deng-Wang Shi thanks Tao Wang  \begin{CJK*}{UTF8}{gbsn}(王涛)\end{CJK*},
and Song-Yao Bai \begin{CJK*}{UTF8}{gbsn}(白松瑶)\end{CJK*}
for helpful discussion. This work is supported by National Natural Science Foundation of China (grant Nos. 11963001
and 12133003).
	
\end{acknowledgements}


\clearpage

\begin{table*}[htbp]
	\centering	
	\caption{The Definitions, the Units, the Prior{, the Medians, $1-\sigma$ Bounds, and the best-fitting values} of the Parameters of the \Ni plus CSI model.}
	\begin{tabular}{llllll}
		\hline
		\hline		
		Parameters	&Definition  & Unit  & Prior   	& Medians & Best-fitting\\
		&   &    &    	&  &  values\\
		\hline
		$M_{\rm ej}$                    & The ejecta mass                                               &   M$_\odot$             &    $[0.1, 50]$    &$12.64^{+0.58}_{-0.55}$ &12.41 \\
		$v_{\rm ph}$                    & {The} early-time photospheric velocity               		    &   {$10^4$ km s$^{-1}$}  &    $[0.1, 5]$     &$1.91^{+0.07}_{-0.07}$ &1.89 \\
		$M_{\rm Ni}$                    & The \Ni mass                                 					&   M$_\odot$          	  &	   $[0, 5]$  	  &$0.41^{+0.00}_{-0.00}$ &0.41\\
		$M_{\rm CSM}$                   & The CSM mass  												&   M$_\odot$             &    $[0.1, 50]$    &$5.87^{+0.26}_{-0.27}$ &5.82\\
		{$\rho_{\rm CSM}$}            & The CSM density 											    &   $10^{-15}$ g cm$^{-3}$&    $[0.01, 100]$  &$3.07^{+0.41}_{-0.42}$ &3.16 \\
		$\Delta t$                      & Time difference value											&   days         		  &    $[11, 15]$	  &$11.89^{+0.27}_{-0.28}$ &11.84 \\
		$\epsilon$                      & The conversion efficiency from the kinetic energy to radiation&                         &    $[0.1, 0.5]$   &$0.28^{+0.03}_{-0.03}$ &0.28 \\
		$x_{\rm 0}$                     & Dimensionless radius                                          &                         &    $[0.01, 0.5]$  &$0.28^{+0.01}_{-0.01}$ &0.28 \\
		$t_{\rm b}$                     & The start time of the linear temperature mode                 &   days                  &    $[10, 30]$     &$11.32^{+0.73}_{-0.71}$ &11.57\\
		$\alpha_{\rm pb}$               & The slope of the temperature curve at the late epochs         &   K d$^{-1}$            &    $[30, 55]$     &$41.52^{+0.64}_{-0.65}$ &41.50 \\
		$t_{\rm shift}$                 & The explosion time relative to the first data                 &   days                  &    $[-20, 0]$     &$-11.85^{+0.25}_{-0.24}$ &$-11.74$\\
		$\chi^{\rm 2}$/dof              &                                                               &                         &                   &114.83  &114.48\\	
		\hline
	\end{tabular}
	\label{table:NI+CSI}
\end{table*}

\clearpage

\begin{table*}[htbp]
	\centering
	\caption{{Comparisons between the Properties of SN~2017dio and Those of Other SNe Ic-BL and HNe}.}
	\begin{tabular}{llllll}
		\hline
		\hline	
		SN name & $E_{\rm K}$ ($10^{52}$ erg) & $M_{\rm ej}$ (M$_\odot$) & $M_{\rm Ni}$ (M$_\odot$) & $v_{\rm ph}$ (km s$^{-1}$)  & References\\
		\hline
		SN 2017dio&2.66   &12.41  & 0.17 &18900    &This paper	 	    \\
		SN 2017ens&1.5 &5& -    & -       &1   \\
		SN 1997ef &0.8     	    &10		 	 &$0.15\pm0.03$	& 20000		     &2	\\
		SN 1998bw &$2-5$        &$8\pm2$     &$0.4-0.7$   	& 30000		     &3		    		\\	
		SN 2003dh & $3.5\pm1.5$ &$7\pm3$  	 &0.4           & 30000$-$40000	 &4	\\	
		SN 2006aj &0.2	    	&2		     &0.2		 	& 20000	         &5	\\	
		Ic-BL(Mean) &1.26 	    &5.42		 &0.36		    & 15114	         &6		\\	
		Ic-BL(Median) &1.09 	   &3.90		&0.26		 & 14000  	     &7		\\
		Ic-BL(Median) &0.51 	   &3.1		    &0.292		 & -		     &8		\\
		Ic-BL(Median \& Mean) & - 	   &-		    &0.15		 & -		 &9		\\
		\hline  				
	\end{tabular}
	\tablecomments{(1)\cite{Chen2018}; (2) \cite{Iwamoto2000}; (3) \cite{Iwamoto1998}; \cite{Nakamura2001}; \cite{Patat2001}; (4) \cite{Woo2003}; \cite{Deng2005} (5) \cite{Mazzali2006}; (6) \cite{Cano2013}; (7) \cite{Cano2013}; (8) \cite{Taddia2019}; (9) \cite{Afsariardchi2021}}
	\label{table:BL-Lc}
\end{table*}

\clearpage

\clearpage

\clearpage

\clearpage
\bibliographystyle{aasjournal}

\begin{appendix}
	
	\newcounter{Afigure}
	\setcounter{Afigure}{1}
	\renewcommand{\thefigure}{A\arabic{Afigure}}
	
	\newcounter{Atable}
	\setcounter{Atable}{1}
	\renewcommand{\thetable}{A\arabic{Atable}}

\begin{figure*}[htbp]
		\centering
		\includegraphics[width=0.24\textwidth,angle=0]{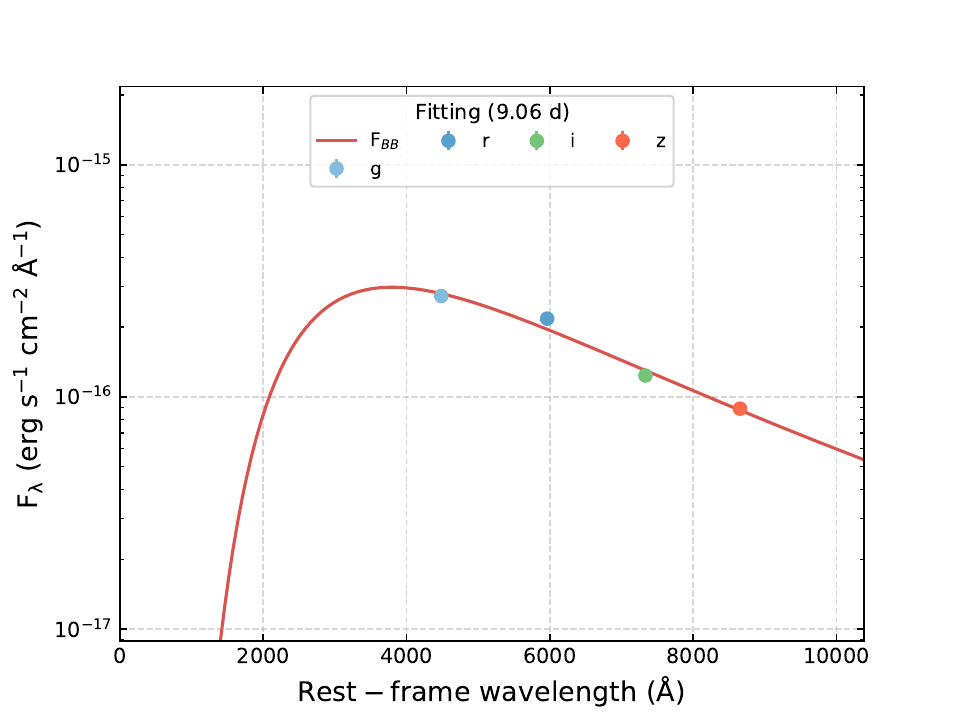}
		\includegraphics[width=0.24\textwidth,angle=0]{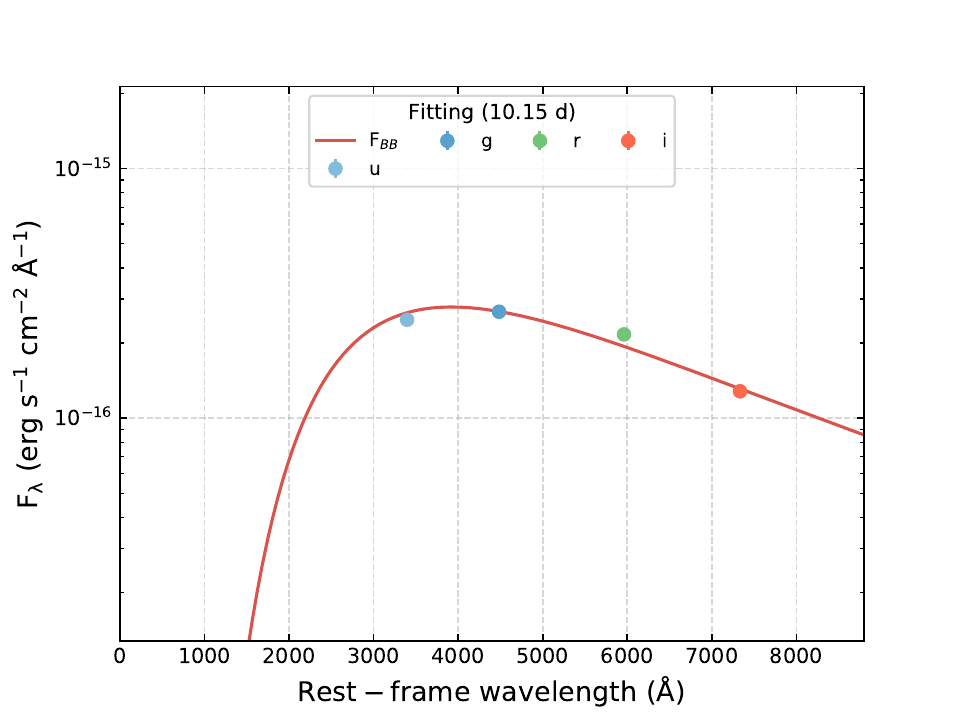}
		\includegraphics[width=0.24\textwidth,angle=0]{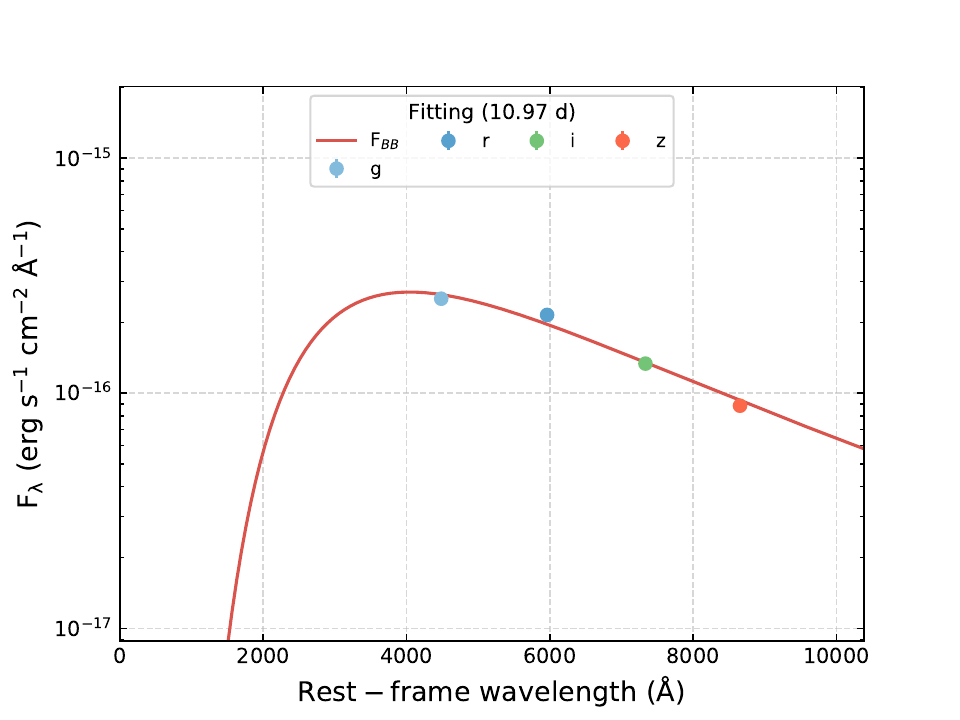}
		\includegraphics[width=0.24\textwidth,angle=0]{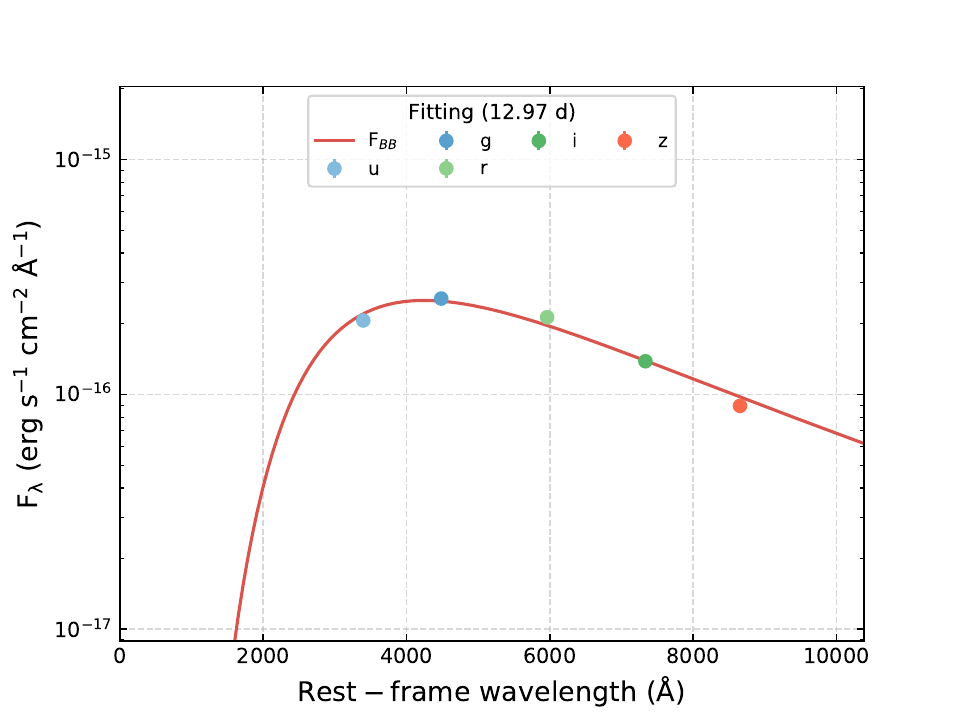}
		\includegraphics[width=0.24\textwidth,angle=0]{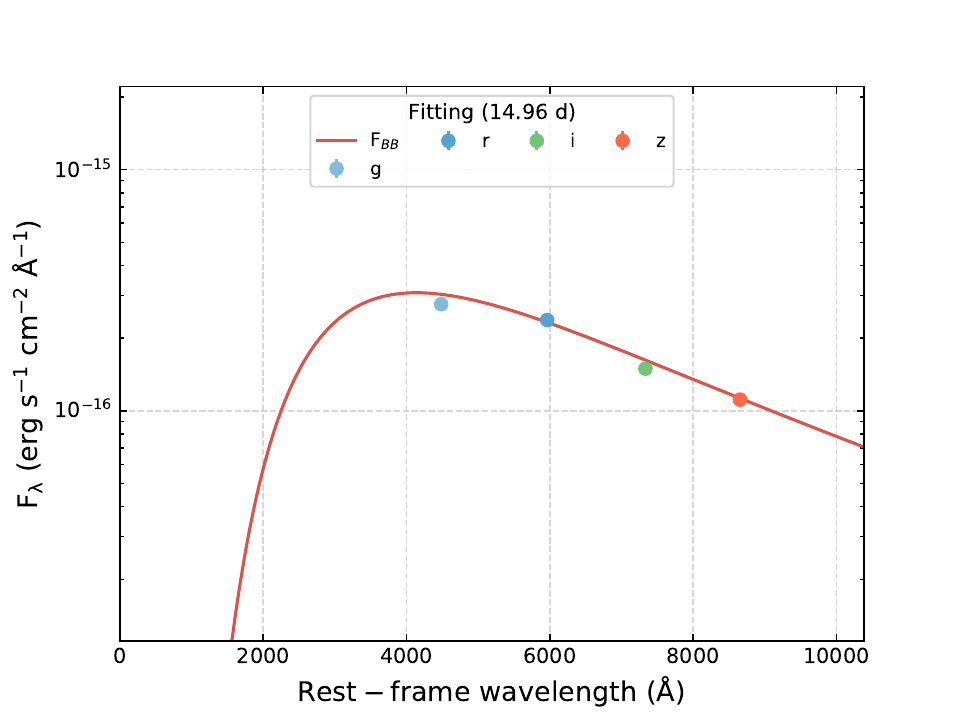}
		\includegraphics[width=0.24\textwidth,angle=0]{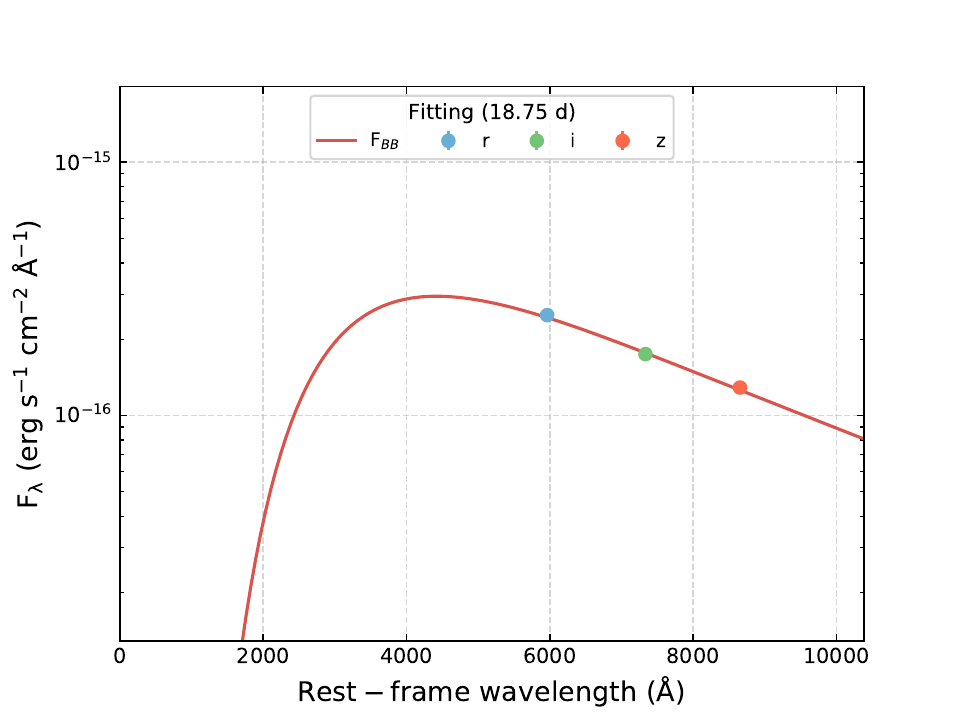}
		\includegraphics[width=0.24\textwidth,angle=0]{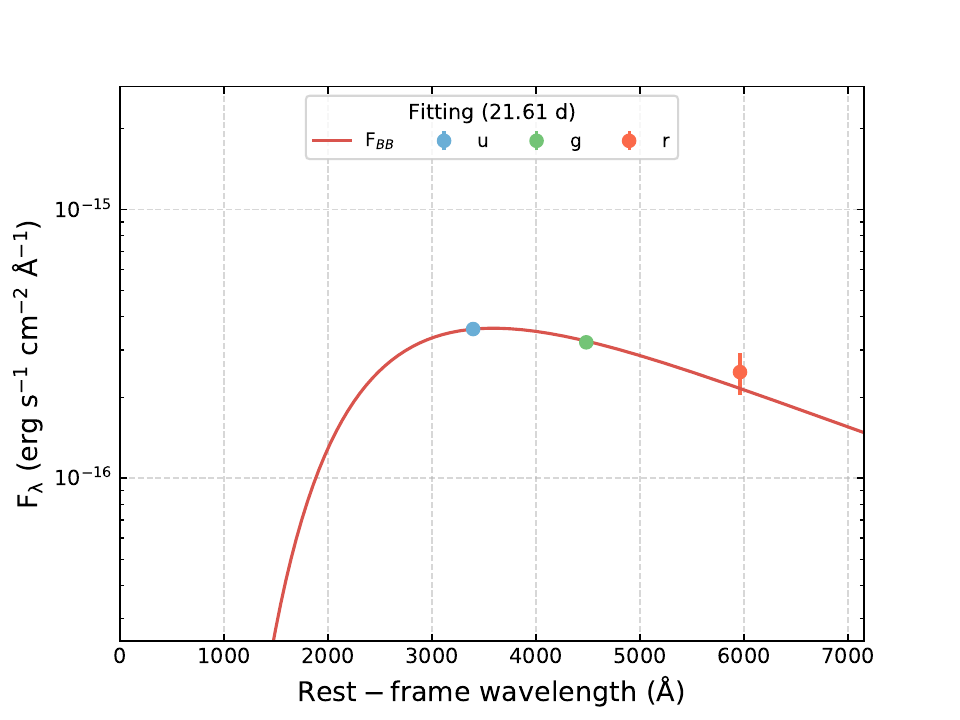}
		\includegraphics[width=0.24\textwidth,angle=0]{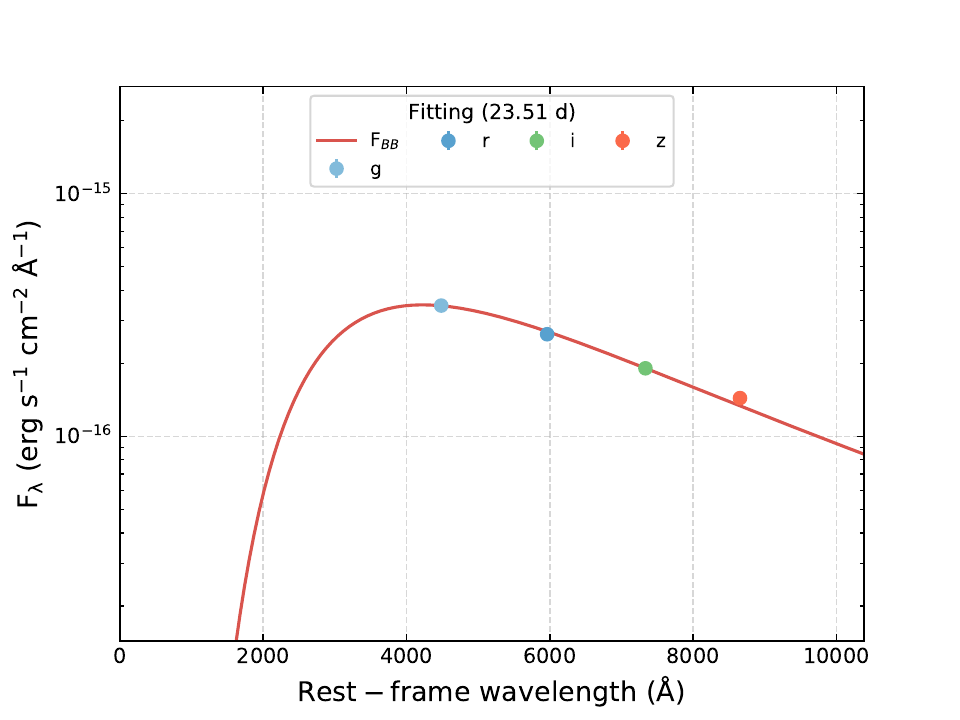}
		\includegraphics[width=0.24\textwidth,angle=0]{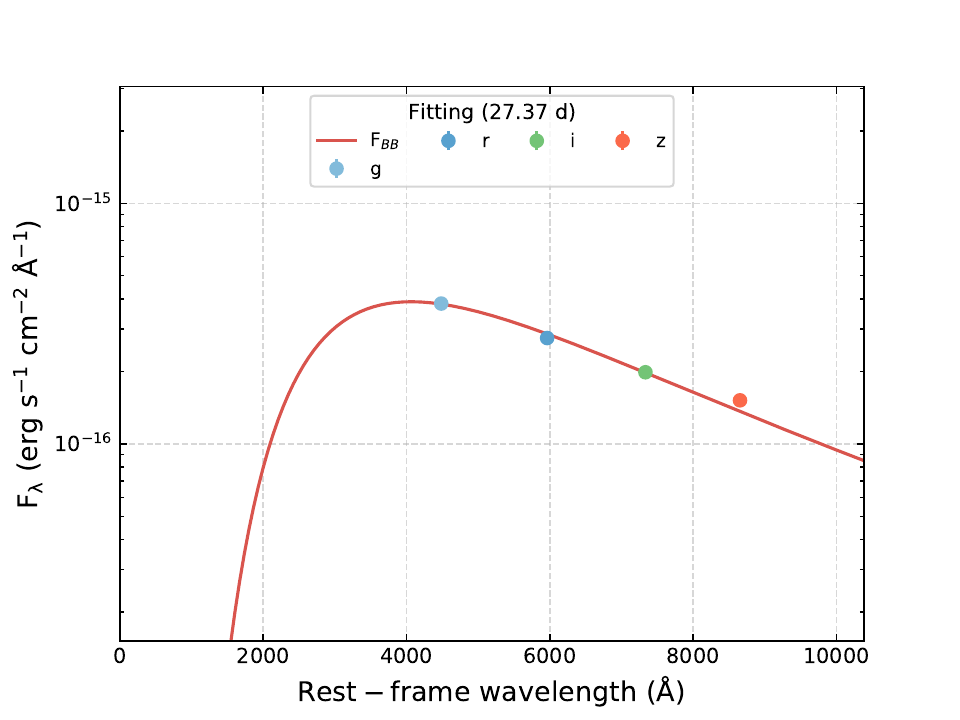}
		\includegraphics[width=0.24\textwidth,angle=0]{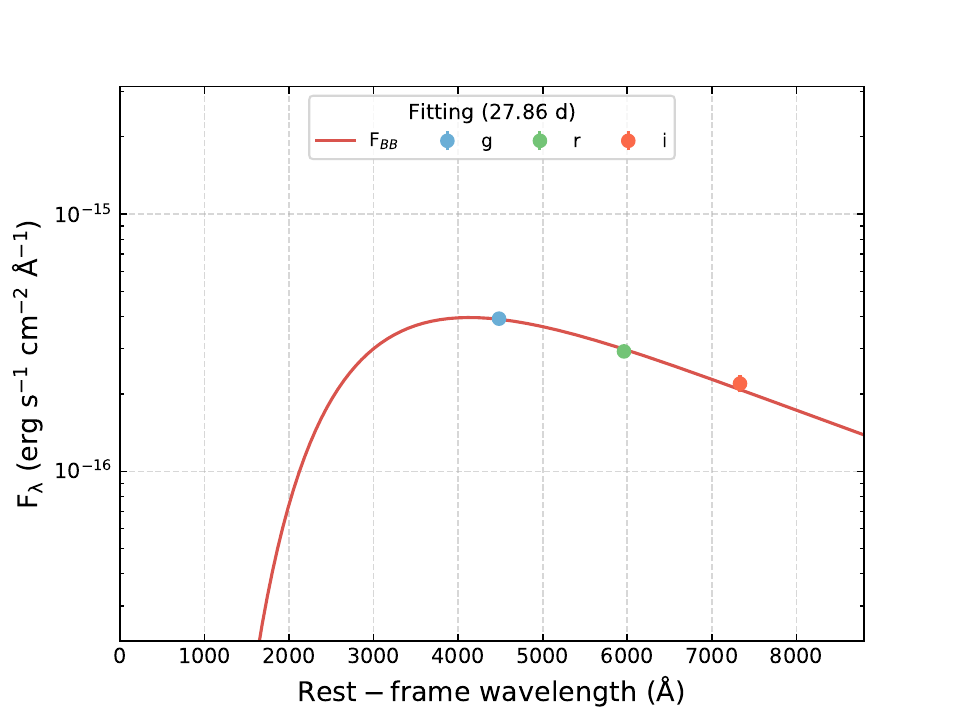}
		\includegraphics[width=0.24\textwidth,angle=0]{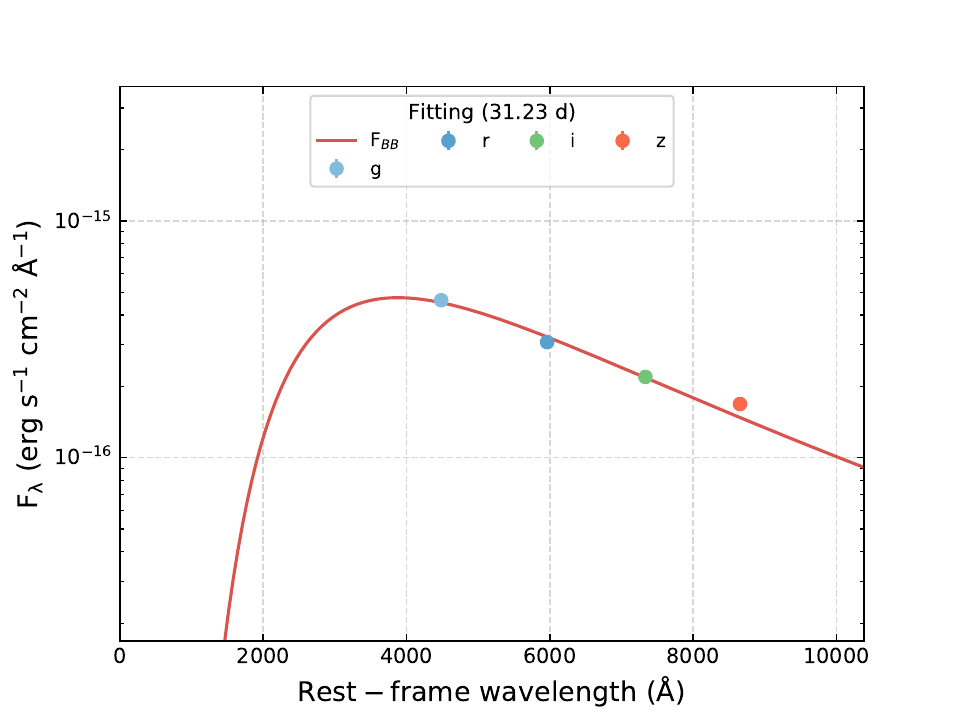}
		\includegraphics[width=0.24\textwidth,angle=0]{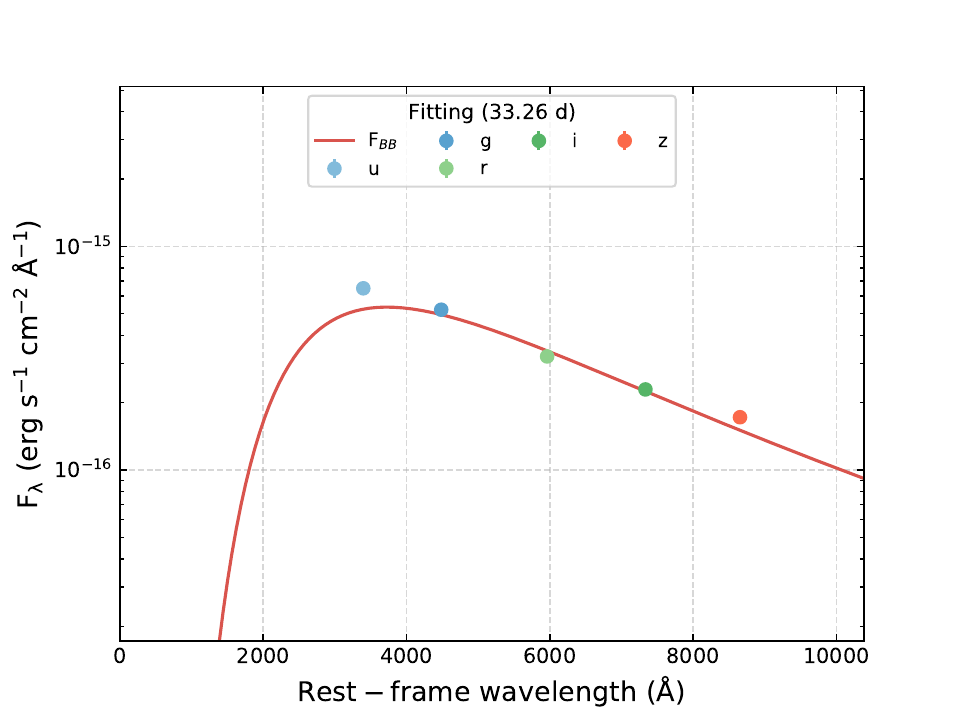}
		\includegraphics[width=0.24\textwidth,angle=0]{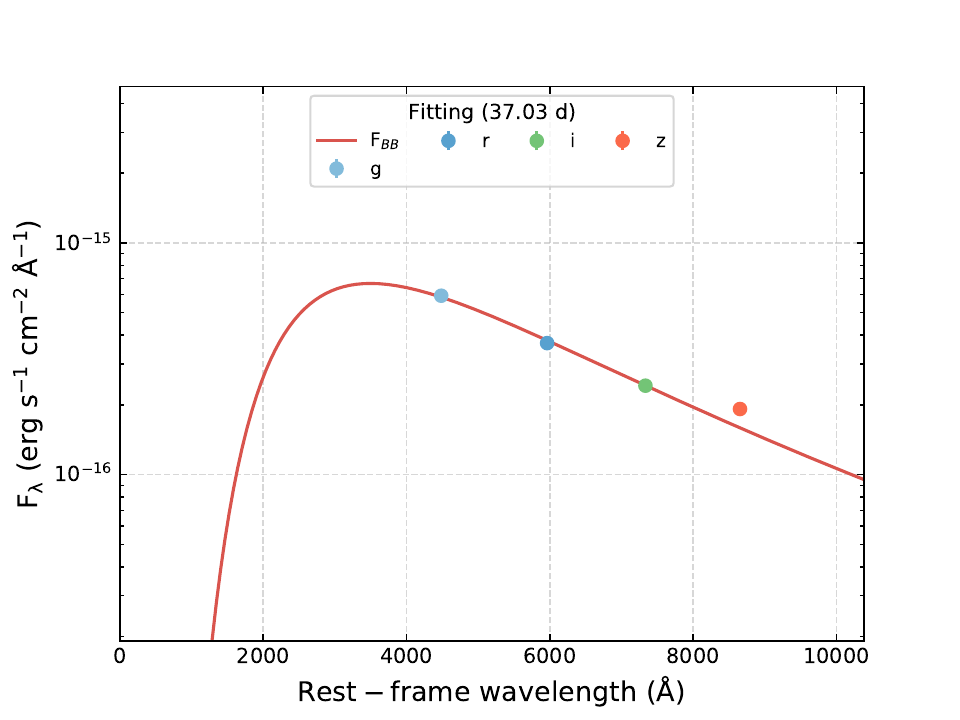}
		\includegraphics[width=0.24\textwidth,angle=0]{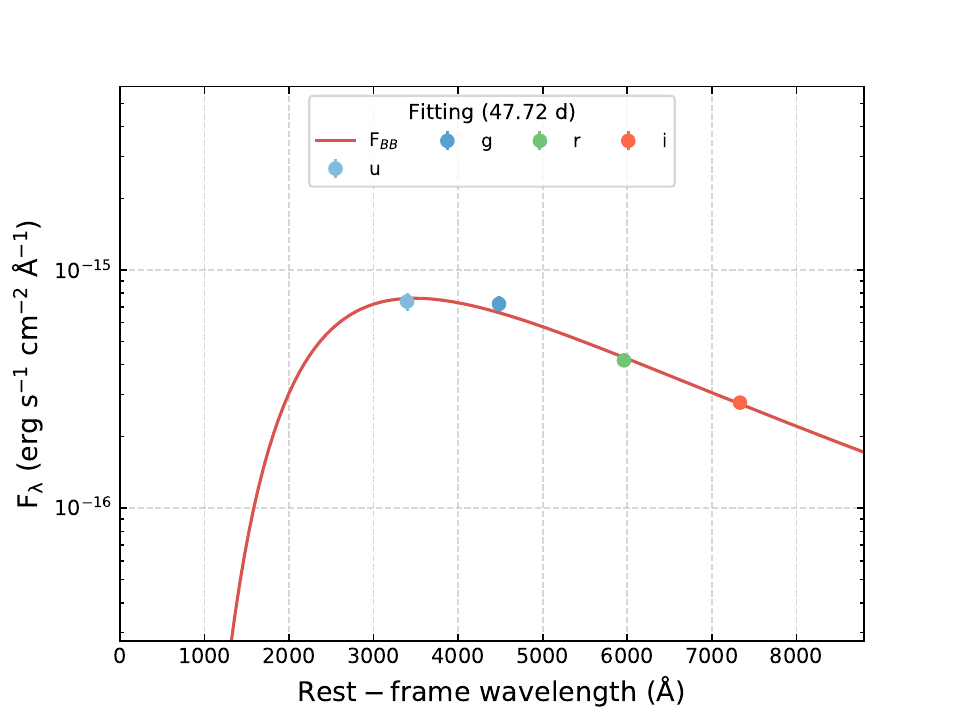}
		\includegraphics[width=0.24\textwidth,angle=0]{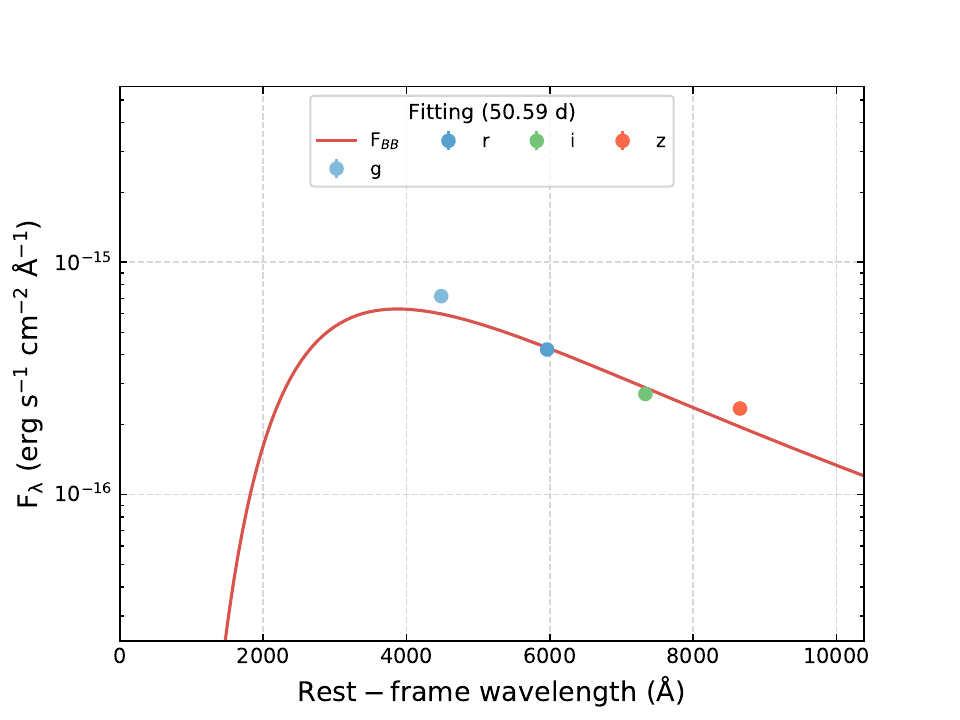}
		\includegraphics[width=0.24\textwidth,angle=0]{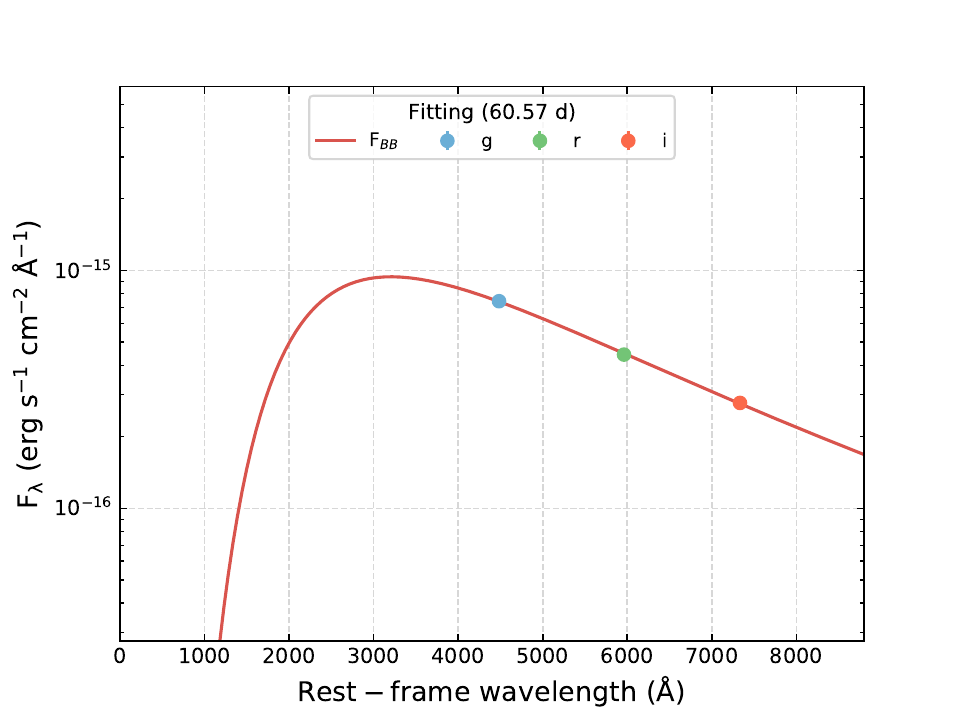}
		\includegraphics[width=0.24\textwidth,angle=0]{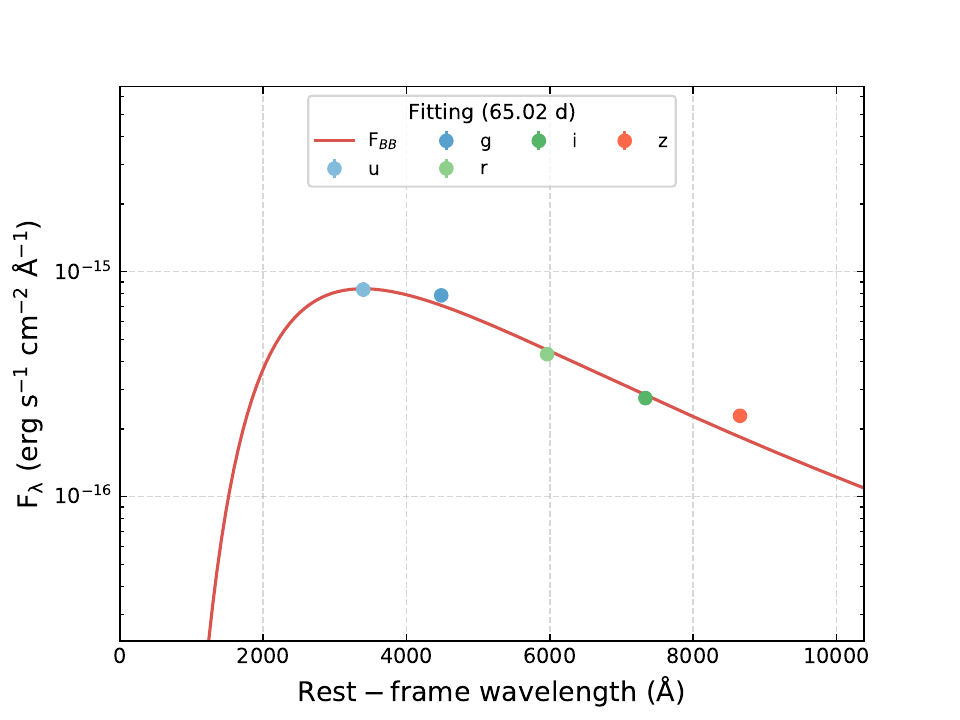}
		\includegraphics[width=0.24\textwidth,angle=0]{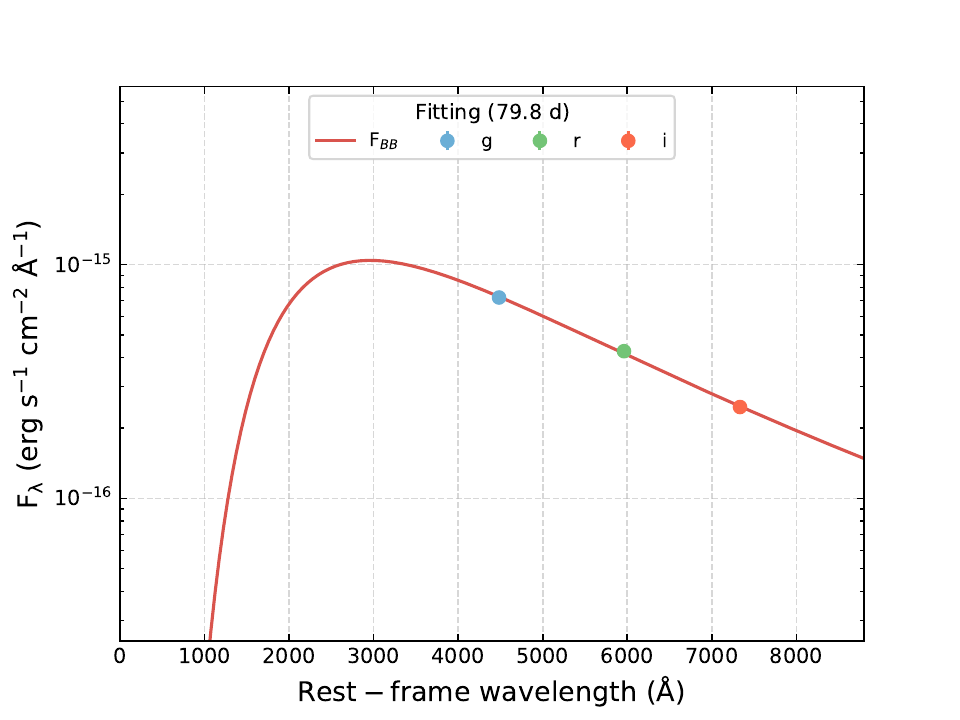}
		\includegraphics[width=0.24\textwidth,angle=0]{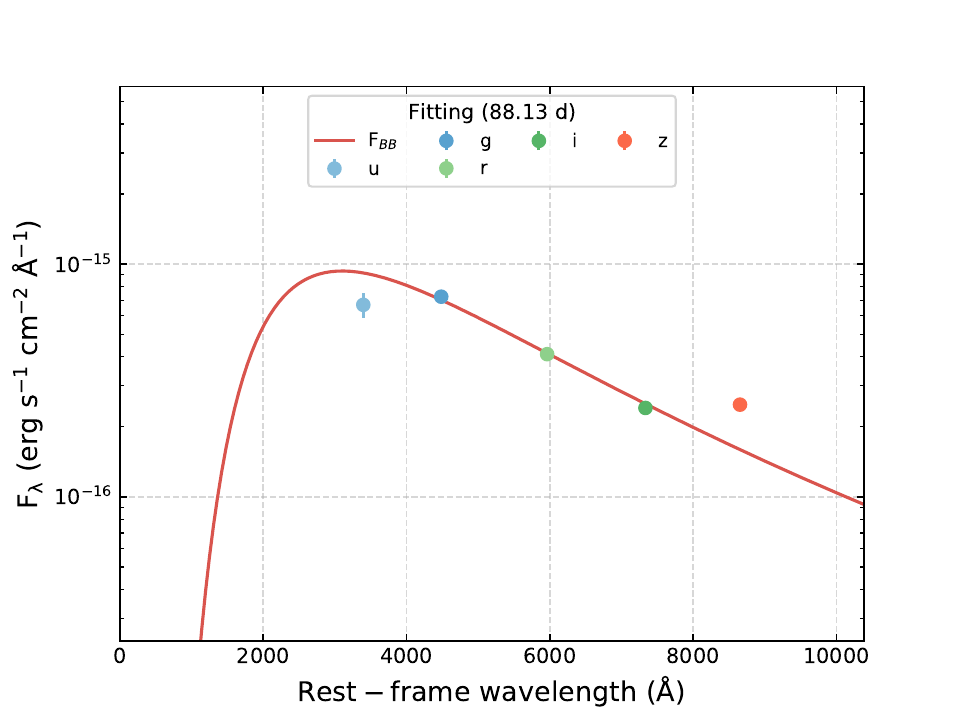}
		\caption{Fits of the blackbody model for SEDs of SN~2017dio. The data (in $ugriz$ bands) are from \cite{Kuncarayakti2018}.}
		\label{fig:sed}
	\end{figure*}

\begin{figure*}[htbp]
		\centering
		\includegraphics[width=0.9\textwidth,angle=0]{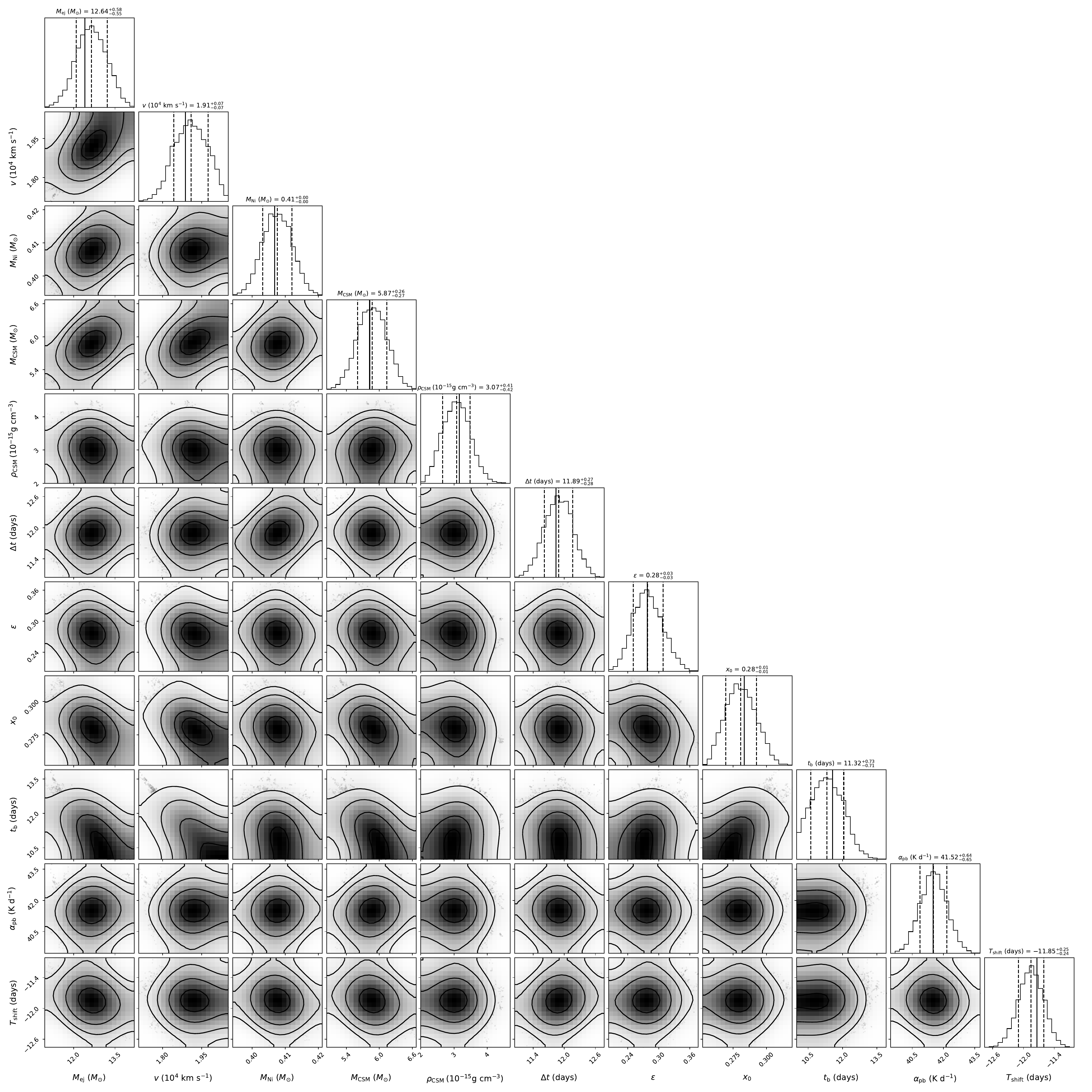}
		\caption{The corner plot of the \Ni plus CSI model. The solid vertical lines represent the best-fitting parameters, while the dashed vertical lines
			represent the medians and the 1-$\sigma$ bounds of the parameters.}
		\label{fig:corner}
	\end{figure*}

	Table \ref{table:blackbody} lists the best-fitting parameters
	(the temperature of the photosphere $T_{\rm ph}$, the radius of the photosphere $R_{\rm ph}$) of the blackbody model for the SEDs of SN~2017dio.
	Figures \ref{fig:sed} and \ref{fig:corner} present the fits of the blackbody model for SEDs of SN~2017dio and the corner plot of the \Ni plus CSI model,
	respectively.
	
	\clearpage
	
	\begin{table*}
		\setlength{\tabcolsep}{18pt}
		\renewcommand{\arraystretch}{1.5}
		\caption{The best-fitting parameters (the temperature of the photosphere $T_{\rm ph}$, the radius of the photosphere $R_{\rm ph}$) of the blackbody model for the SEDs of SN~2017dio.}
		\label{table:blackbody}
		\centering
		\begin{tabular}{cccc}
			\hline\hline
			Phase\footnote{All the phases are relative to the date of the first data which is MJD 57865.5. All epochs are transformed to the rest-frame ones.}
			& $T_{\rm ph}$&$R_{\rm ph}$ & $\chi^{\rm 2}$/dof \\
			(days) & (10$^{3}$\,K) & (10$^{15}$ cm)    & \\\hline
			9.06 	&$	7633.33 	^{+	73.34 	}_{-	71.42 	}$&$	1.55 	^{+	0.02 	}_{-	0.02 	}$&	38.23 	\\\hline
			10.15 	&$	7389.68 	^{+	106.35 	}_{-	106.45 	}$&$	1.62 	^{+	0.04 	}_{-	0.04 	}$&	10.36 	\\\hline
			10.97 	&$	7165.85 	^{+	43.29 	}_{-	42.28 	}$&$	1.73 	^{+	0.02 	}_{-	0.02 	}$&	78.24 	\\\hline
			12.97 	&$	6821.84 	^{+	40.75 	}_{-	40.28 	}$&$	1.88 	^{+	0.02 	}_{-	0.02 	}$&	26.86 	\\\hline
			14.96 	&$	7005.53 	^{+	45.82 	}_{-	46.33 	}$&$	1.95 	^{+	0.02 	}_{-	0.02 	}$&	73.00 	\\\hline
			18.75 	&$	6557.80 	^{+	106.36 	}_{-	100.16 	}$&$	2.25 	^{+	0.06 	}_{-	0.06 	}$&	13.53 	\\\hline
			21.61 	&$	8070.59 	^{+	355.32 	}_{-	319.07 	}$&$	1.48 	^{+	0.16 	}_{-	0.15 	}$&	0.59 	\\\hline
			23.51 	&$	6858.63 	^{+	7.57 	}_{-	7.43 	}$&$	2.19 	^{+	0.01 	}_{-	0.01 	}$&	131.08 	\\\hline
			27.37 	&$	7131.11 	^{+	26.40 	}_{-	25.98 	}$&$	2.10 	^{+	0.01 	}_{-	0.01 	}$&	17.22 	\\\hline
			27.86 	&$	7011.20 	^{+	305.16 	}_{-	275.36 	}$&$	2.21 	^{+	0.19 	}_{-	0.18 	}$&	0.72 	\\\hline
			31.23 	&$	7471.97 	^{+	26.49 	}_{-	26.44 	}$&$	2.06 	^{+	0.01 	}_{-	0.01 	}$&	127.57 	\\\hline
			33.26 	&$	7778.94 	^{+	42.55 	}_{-	41.75 	}$&$	1.98 	^{+	0.02 	}_{-	0.02 	}$&	129.67 	\\\hline
			37.03 	&$	8287.03 	^{+	25.24 	}_{-	24.83 	}$&$	1.89 	^{+	0.01 	}_{-	0.01 	}$&	176.79 	\\\hline
			47.72 	&$	8313.38 	^{+	307.11 	}_{-	297.76 	}$&$	2.00 	^{+	0.12 	}_{-	0.12 	}$&	0.83 	\\\hline
			50.59 	&$	7480.96 	^{+	48.51 	}_{-	48.15 	}$&$	2.37 	^{+	0.02 	}_{-	0.02 	}$&	737.77 	\\\hline
			60.57 	&$	9037.63 	^{+	271.67 	}_{-	253.90 	}$&$	1.80 	^{+	0.08 	}_{-	0.08 	}$&	0.29 	\\\hline
			65.02 	&$	8545.53 	^{+	109.99 	}_{-	108.24 	}$&$	1.96 	^{+	0.04 	}_{-	0.03 	}$&	82.34 	\\\hline
			79.80 	&$	9787.94 	^{+	358.76 	}_{-	339.79 	}$&$	1.56 	^{+	0.08 	}_{-	0.07 	}$&	0.66 	\\\hline
			88.13 	&$	9338.99 	^{+	137.02 	}_{-	130.48 	}$&$	1.66 	^{+	0.03 	}_{-	0.03 	}$&	66.15 	\\\hline
		\end{tabular}
		
	\end{table*}
	
	\clearpage
	
	\clearpage
	\stepcounter{Afigure}
\end{appendix}

\end{document}